\documentclass[
aip,
pop,
amsmath,amssymb,
reprint,%
]{revtex4-1}

\usepackage{graphicx}
\usepackage{dcolumn}
\usepackage{bm}

\usepackage[utf8]{inputenc}
\usepackage[T1]{fontenc}
\usepackage{mathptmx}

\usepackage{color}
\usepackage{amssymb}
\usepackage{natbib}
\usepackage{hyperref}

\begin{document}
	
	
	\title[]{Hybrid simulation of Alfv\'en wave parametric decay instability in a laboratory relevant plasma}
	
	\author{Feiyu Li}
	\email{fyli.acad@gmail.com}
	\affiliation{New Mexico Consortium, Los Alamos, NM 87544, USA
	}
	\author{Xiangrong Fu}
	\affiliation{New Mexico Consortium, Los Alamos, NM 87544, USA
	}
	\author{Seth Dorfman}
	\affiliation{Space Science Institute, Boulder, CO 80301, USA
	}
	\affiliation{University of California Los Angeles, Los Angeles, CA 90095, USA}
	
	\date{\today}
	
	\begin{abstract}
		Large-amplitude Alfv\'en waves are subject to parametric decays which can have important consequences in space, astrophysical, and fusion plasmas. Though the Alfv\'en wave parametric decay instability was predicted decades ago, observational evidence is limited, stimulating considerable interest in laboratory demonstration of the instability and associated numerical modeling. Here, we report  novel hybrid simulation of the Alfv\'en wave parametric decay instability in a laboratory relevant plasma (based on the Large Plasma Device), using realistic wave injection and wave-plasma parameters. Considering only collisionless damping, we identify the threshold Alfv\'en wave amplitudes and frequencies required for triggering the instability in the bounded plasma. These threshold behaviors are corroborated by simple theoretical considerations. Other effects not included in the present model such as finite transverse scale and ion-neutral collision are briefly discussed. These hybrid simulations demonstrate a promising tool for investigating laboratory Alfv\'en wave dynamics that can provide guidance for future laboratory demonstration of the parametric decay instability. 
	\end{abstract}
	
	\maketitle

	\section{Introduction} \label{introduction}
	
	Predicted by Hannes Alfvén~\cite{alfven1942existence} in 1942, the Alfvén wave (AW) is widely conceived to be the fundamental mode of a magnetized plasma~\cite{wilcox1960experimental,hasegawa1982alfven,gekelman1999review}. Because of little geometrical attenuation, shear AWs are prevalent in space and astrophysical plasmas as a carrier of magnetic and flow energy over a large range of scales~\cite{de2007chromospheric}. In laboratory plasmas, AWs also exist in tokamaks and linear devices, and their interaction with energetic particles is crucial to the performance of fusion reactions~\cite{chen2016physics}. Large-amplitude AWs play an important role in several nonlinear processes, such as energy cascade~\cite{kraichnan1965inertial} and particle acceleration~\cite{kletzing1994electron}, which are essential in plasmas mentioned above. Parametric instabilities are an important class of such nonlinear interactions, arising from parametric excitation of collective plasma modes. These instabilities could potentially contribute to coronal heating~\cite{de2007chromospheric,velli1997alfven}, the observed spectrum and cross-helicity of solar wind turbulence~\cite{inhester1990drift,del2001parametric,yoon2008parallel}, and damping of fast magnetosonic waves in fusion plasmas~\cite{lee1998internal,oosako2009parametric}.
	
	Three major types of parametric instabilities driven by a parallel (perpendicular wave vector $k_\perp=0$) AW have been found theoretically~\cite{sagdeev1969nonlinear,hasegawa1976parametric,derby1978modulational,goldstein1978instability,wong1986parametric,longtin1986modulation,hollweg1993modulational,hollweg1994beat}: modulational, beat, and decay instabilities. The modulational instability drives forward upper and lower Alfvénic sidebands, as well as a non-resonant acoustic mode at the sideband separation frequency. To allow the forward AWs to interact, the pump wave must be dispersive, i.e., requiring the finite frequency effect through inclusion of the Hall term~\cite{hollweg1994beat}. The beat instability drives a forward upper Alfvénic sideband and a backward lower Alfvénic sideband and is generally negligible in the low-beta plasmas to be studied here. In contrast, the parametric decay instability (PDI), involving the decay of a forward pump AW into a backward daughter AW and a forward ion acoustic wave, is best known for its prominence in low-beta plasmas. 
	
	The PDI is of special interest to the space plasma community. Firstly, the backward AW generated during the PDI can interact with the forward AW, which is an important ingredient for developing magnetohydrodynamics (MHD) turbulence~\cite{van2011heating}. Secondly, the PDI-produced ion acoustic wave can strongly heat ions through wave damping~\cite{fu2020heating}. Recently, Fu et al.~\cite{fu2018parametric} showed that the PDI and associated kinetic heating can be robust even in a turbulent environment, which is often the case in space plasmas. The PDI has a low-beta favored growth rate, $\gamma_g\propto \beta^{-1/4}$, according to a linear analysis~\cite{sagdeev1969nonlinear}. A vast region of the solar surface extending from the chromosphere to corona is dictated by low-beta plasmas~\cite{gary2001plasma}, and several studies have suggested that the PDI play important roles in a stratified chromosphere~\cite{matsumoto2014connecting}, in an expanding solar wind~\cite{shoda2018frequency,reville2018parametric}, and in the solar wind near 1 AU~\cite{shi2017parametric,bowen2018density}, contributing to turbulence development and plasma heating.
	
	The PDI of a shear AW was first predicted by Sagdeev and Galeev~\cite{sagdeev1969nonlinear} in 1969 and analyzed in the low-amplitude low-beta limit for a linearly polarized AW. Derby~\cite{derby1978modulational} and Goldstein~\cite{goldstein1978instability} later independently extended the analyses to finite wave amplitudes for circular polarization in the single-fluid MHD framework. The dispersive effects arising from ion cyclotron resonance were further considered with a two-fluid framework~\cite{wong1986parametric,hollweg1994beat}. Despite the firm theoretical bases, observational evidence of PDI in space has been scarce. The satellite observations in the upstream of the bow shock of Earth’s magnetosphere may have found a number of cases with possible AW decay signatures~\cite{spangler1997observations,narita2007npgeo,dorfman2017foreshock}. A statistical study~\cite{bowen2018density} has shown correlations of density or magnetic fluctuations with PDI over a broad parameter space. However, unambiguous observation of individual PDI events in space requires overcoming several challenges~\cite{narita2007npgeo}, such as the turbulent environment, limited sampling locations, broad pump bandwidths, and various kinetic effects~\cite{spangler1997observations}. Therefore, there is a need for laboratory experiments to validate theoretical predictions. Particularly, the active control of laboratory wave and plasma conditions may help to elucidate the underlying physical processes in a manner not possible with space measurements. 
	
	The long wavelengths of AWs makes it difficult to fit the waves in a laboratory device while maintaining a low enough plasma density to allow for the use of physical probes throughout the plasma region. Recently, with the Large Plasma Device (LAPD) at University of California, Los Angeles (UCLA), several controlled AW experiments have been conducted~\cite{gekelman1997laboratory,gekelman1999review,maggs2003laboratory,vincena2006drift,auerbach2010control,auerbach2011resonant,howes2012toward,dorfman2013nonlinear,dorfman2016observation}. Particularly, Dorfman and Carter~\cite{dorfman2013nonlinear} excited an ion acoustic mode through beating of two counter-propagating AWs. While this setup is similar to PDI excitation in the seeded mode (i.e., seeding the PDI with an artificial daughter AW), the instability growth was not clearly identified. More recently, Dorfman and Carter~\cite{dorfman2016observation} also recorded a perpendicular modulational instability of kinetic AWs; when a single finite-frequency, finite-$k_\perp$, AW was launched above a certain wave amplitude threshold, three daughter waves were detected: two sideband AWs co-propagating with the pump wave and a low frequency non-resonant mode at the sideband separation frequency. However, the theoretical growth rate for zero-$k_\perp$ modulational instability was too small to explain the observation. More critically, the PDI was missing despite having a much larger theoretical growth rate than the zero-$k_\perp$ modulational instability under the conditions investigated. So far, no quantitative evidence of PDI has been found in laboratory experiments. In planning future experiments toward demonstration of this important instability, it should be of considerable interest to find optimal parameters with first-principle numerical simulations. 
	
	In this paper, we report the development of novel hybrid simulation capabilities aiming at direct modeling of laboratory AW dynamics. In particular, we simulate the PDI of nonlinear AWs with physical setups and wave-plasma parameters that closely resemble LAPD conditions. As an important step, we consider LAPD-like injection of a plane AW in a bounded (along the background magnetic field direction) plasma with ion kinetics retained (i.e., collisionless Landau damping). These simulations allow us to gain insight into the threshold AW amplitudes and frequencies needed for triggering PDI in a collisionless bounded plasma, and the results can be interpreted with simple theoretical estimates. We also briefly discuss complicating effects such as ion-neutral collisions and finite AW source sizes not included in the current model. It is hoped that the present work will further developments in hybrid simulation of realistic laboratory plasmas and provide guidance for future demonstration of PDI in LAPD.

	\section{Hybrid simulation of PDI with LAPD-like wave injection and wave-plasma parameters} \label{setup}
	
	\begin{figure*}[htp]
		\centering
		\includegraphics[width=0.9\textwidth]{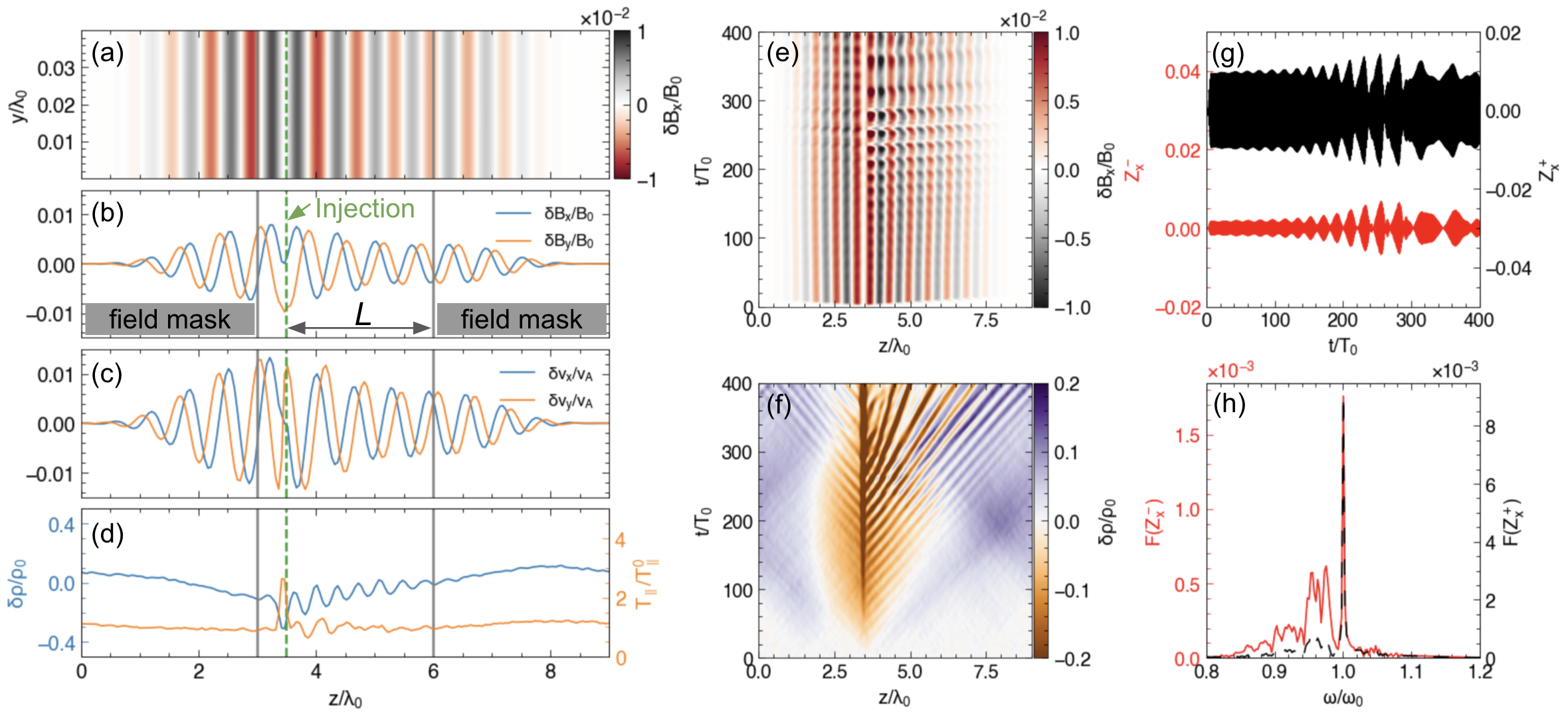}
		\caption{Hybrid simulation of Alfv\'en wave parametric decay instability using LAPD-like wave injection and wave-plasma parameters. Snapshots of (a) 2D wave pattern and $y$-averaged (b) wave magnetic fields, (c) velocity fields, and (d) density fluctuation and parallel ion temperature at $t=200T_0$. (e, f) Spatiotemporal evolution of the $B_x$ and density fluctuation, respectively. Notice the AW looks like immobile due to a resolution issue; $B_x$ data are dumped about every wave period. (g) Wave signatures probed at $z=4.58\lambda_0$ and (h) corresponding Fourier spectra, where the black and red curves refer to forward and backward going waves, respectively.}
		\label{fig1-hybrid}
	\end{figure*}

	We start with introducing our hybrid simulation model. The PDI has been extensively studied via MHD simulations~\cite{del2001parametric,shi2017parametric}, hybrid simulations~\cite{terasawa1986decay,vasquez1995simulation,araneda2007collisionless,matteini2010parametric,fu2018parametric,gonzalez2020role}, and even full particle-in-cell simulations~\cite{nariyuki2008parametric}. In the context of Earth’s foreshock and solar wind, the PDI has also been invoked in simulations to explain, for example, the generation of density fluctuations~\cite{tanaka2007parametric}, the origin of low-frequency Alfvénic spectrum~\cite{reville2018parametric}, the effects of wind acceleration or expansion~\cite{shoda2018frequency}, and the evolution of magnetic-field-line switchbacks~\cite{tenerani2020magnetic}. However, the above-mentioned simulations have mostly considered a periodic system~\cite{shi2017parametric,nariyuki2008parametric,tenerani2020magnetic,gonzalez2020role}, limiting their applicability to laboratory plasmas where the interactions are finite in space-time and nonperiodic. The periodic boundary constraint was lifted in some MHD simulations~\cite{del2001parametric}, but they did not capture kinetic effects which are important in laboratory PDI. To address that, we have recently developed novel hybrid simulation capabilities toward direct modeling of laboratory AW dynamics. 
	
	The new developments are based on the three-dimensional hybrid code H3D ~\cite{karimabadi2006global}, which treats electrons as a massless fluid and ions as individual macroparticles. Therefore, the simulation captures ion kinetic effects and also saves computation by ignoring electron dynamics. The code was modified to simulate turbulent plasmas with periodic boundary conditions~\citep{fu2018parametric,fu2020heating}. Recently, we have relaxed the periodic boundary constraint and studied the PDI in a large system ($\sim 100\lambda_0$ with $\lambda_0$ being the reference AW wavelength) of absorption boundaries for the AWs~\cite{li2022parametric}. Compared to usual periodic interaction, several new PDI dynamics were found with the nonperiodic interaction, including reduced energy transfer, and localized density fluctuations and ion heating, as well as the formation of a stable residual wave packet that is not affected by PDI. Though the parameters (e.g., $\delta B/B_0\sim 0.1, \beta\sim 0.01$) used were still away from typical LAPD conditions (e.g., $\delta B/B_0\sim 10^{-3}, \beta\sim 10^{-4}-10^{-3}$), the study uncovered several limitations of usual periodic boundary condition in addressing AW dynamics in a laboratory setting.

	In the present work, we further implement wave injection at prescribed locations, similar to the wave injection from an antenna in LAPD experiments, and investigate the resultant PDI dynamics. The basic simulation setup is illustrated in Fig.~\ref{fig1-hybrid}. A typical run involves one grid cell for the $x$ direction and four cells for the $y$ direction. The $z$ direction (i.e., the AW propagation direction) has a size of $9\lambda_0$, corresponding to $90d_i$, where $d_i=c/\omega_{pi}$  is the ion inertial length, $c$ is the light speed in a vacuum, and $\omega_{pi}$ is the ion plasma frequency. The cell sizes are $\Delta x=1 d_i$, $\Delta y=0.1d_i$, $\Delta z=0.5d_i$, and the time step size is $\Delta t=0.01 \Omega_{ci}^{-1}$, where $\Omega_{ci}$ is the ion cyclotron frequency and satisfies $\omega_{pi}/\Omega_{ci}=c/v_A=300$ with $v_A$ being the Alfv\'en speed. The ions are sampled by 1000 macroparticles per cell. The ion-to-electron temperature ratio, $T_i/T_e$, can be varied to approximate different LAPD discharge conditions. A finite resistivity of $\eta=2\times10^{-4}$ is used to model the spatial wave damping as found in LAPD~\cite{gigliotti2009generation}. The wave injection can be prescribed at any locations within the domain and typically involves an up-ramp of $5T_0$ in temporal profile followed by a long plateau (100s $T_0$), where $T_0=\lambda_0/v_A$.  
	
	Our simulations essentially model the injection of a plane-wave with $k_\perp=0$. In this quasi-1D setup, we do not consider the finite transverse scale and large $k_\perp$ as found in actual experiments~\cite{gigliotti2009generation,karavaev2011generation}. Moreover, the ion-neutral collisions present in a partially ionized LAPD plasma~\cite{dorfman2013nonlinear} are not modelled in the current code. 
	Despite the simplifications, our simulations and associated analyses should be of practical interest as we adopt realistic wave and plasma parameters (e.g., wave amplitude, polarization, and plasma beta, temperature) and geometry (e.g., plasma size, wave injection, and boundary conditions). Implementation of these features is an important step toward fully modeling LAPD experiments. The wave absorption at boundaries is achieved using field masks as reported in our previous work~\cite{li2022parametric}.
	Notice that the periodic boundary is still used for ion particles. To reduce the effects of particle circulation in the simulation domain, we use the field masks as a buffer area for particles, as illustrated by the shaded areas in Fig.~\ref{fig1-hybrid}(b). For typical runs, the buffer area is as large as the central unmasked region, such that the particle circulation effect on wave-plasma interactions in the central region is negligible. 
	
	Figure~\ref{fig1-hybrid} presents a case where strong PDI is observed. A left-hand polarized AW of normalized frequency $\omega_0/\Omega_{ci}=0.63$ is injected at  $z=3.5\lambda_0$ (green dashed line) by prescribing its magnetic field perturbations as
	\begin{equation}
		\frac{\delta \vec{B}}{B_0} =  \left[\frac{\delta B_x}{B_0}\sin(\omega_0t)\hat{x} 
		+ \frac{\delta B_y}{B_0}\cos(\omega_0t)\hat{y}\right]\chi(t),
		\label{eq:init_wave}
	\end{equation}
	where $\frac{\delta B_x}{B_0}=\frac{\delta B_y}{B_0}=\frac{\delta B}{B_0}$ and $\chi(t)$ is the temporal profile. The velocity field perturbations, $\delta v$, are automatically generated by the field solver. The finite-$\frac{\omega_0}{\Omega_{ci}}$, zero-$k_\perp$ planar AW is verified to follow the dispersion~\cite{gekelman1997laboratory,stasiewicz2000small} $\frac{\omega_0}{k_\parallel}=v_A\sqrt{1-(\frac{\omega_0}{\Omega_{ci}})^2}$ and has unequal magnetic/velocity perturbations with the ratio~\cite{hollweg1999kinetic} $R_{bv}\equiv\frac{\delta B/B_0}{\delta v/v_A}=\frac{\omega_0}{k_\parallel v_A}=\sqrt{1-(\frac{\omega_0}{\Omega_{ci}})^2}$. The wave amplitude of the present case, $\delta B/B_0=0.01$, is  higher than normally achievable with LAPD, but may be realized through, for example, tuning the background magnetic field $B_0$ or AW maser~\cite{maggs2003laboratory}. The wave pattern is shown in Fig.~\ref{fig1-hybrid}(a-c) at $t=200T_0$, when the PDI has sufficiently developed; the line plots are averaged over the $y$ axis, with which the results have little dependence because of the plane-wave injection. While the wave specified by Eq.~(\ref{eq:init_wave}) propagates along the $+z$ direction, the field solver also generates a mirror wave of the same polarization and amplitude that propagates in the $-z$ direction. These waves start to be damped when they reach the mask regions on both sides. As such, the effective plasma length that the forward wave travels through is $L=2.5\lambda_0=25d_i$ as marked in Fig.~\ref{fig1-hybrid}(b); this length is typical for LAPD experiments\citep[e.g.][]{dorfman2016observation}. 
	
	\begin{figure*}[t]
		\centering
		\includegraphics[width=0.9\textwidth]{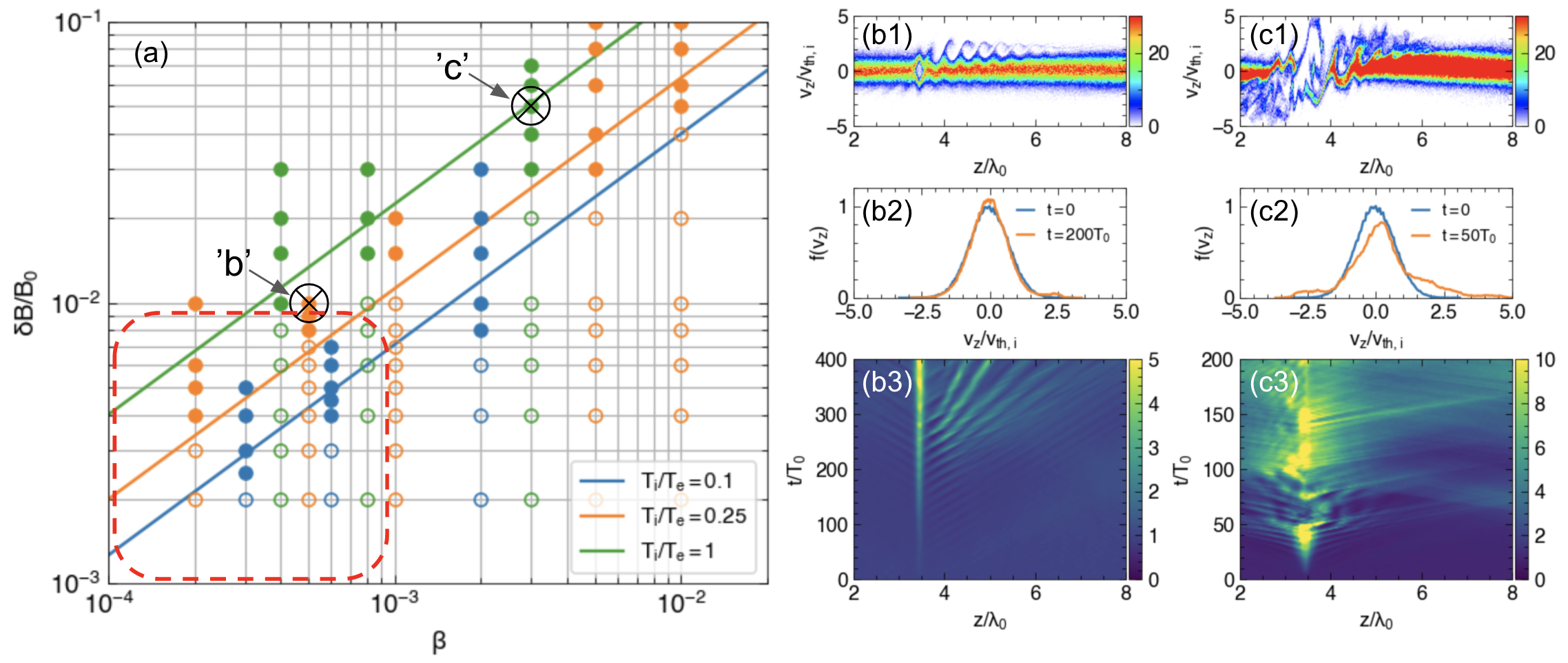}
		\caption{Threshold Alfv\'en wave amplitude for triggering the parametric decay instability. (a) The parameter space spanned by wave amplitude and plasma beta. Typical LAPD conditions are indicated by the dashed square area in the bottom-left corner. Each dot represents a hybrid simulation. The open dots indicate no PDI found and the filled dots indicate the opposite. Different colors refer to different ion-to-electron temperature ratios. The solid lines are theoretical predictions. (b1-b3) show the ion phasespace distribution $z-v_z$, velocity distribution, and spatiotemporal evolution of ion parallel temperature (normalized to initial temperature), respectively, for the case marked as orange cross in (a). The velocity distributions show two instants at $t=0, 200T_0$, respectively, for ions within $z=[3.5, 5]\lambda_0$ only. (c1-c3) corresponds to the green-cross case marked in (a).}
		\label{fig2-amp}
	\end{figure*}

	One signature of PDI is the excitation of an ion acoustic wave (e.g., displayed as density fluctuations) and associated ion heating, as seen in Fig.~\ref{fig1-hybrid}(d). To better illustrate the whole dynamics, Figure~\ref{fig1-hybrid}(f) shows the space-time evolution of $\delta\rho/\rho_0$, where the density fluctuations emerge near the wave injection point in just 10s $T_0$ and then gradually propagate outward. The density modulation deepens with time, an indication of instability growth. A deep narrow density trough is gradually formed at $z=3.5\lambda_0$ due to continuous wave injection and localized plasma heating [e.g., the orange curve in Fig.~\ref{fig1-hybrid}(d)]. The density fluctuations propagates forward in the region $z>3.5\lambda_0$, consistent with the acoustic wave direction of PDI driven by a forward pump wave. At later times ($t\gtrapprox 300T_0$), the acoustic wave even forms nonlinear structures propagating at a reduced speed, due to ion trapping (not shown). In PDI, the large density fluctuations further beat with the pump wave, generating a backward daughter AW; the latter overlaps with the pump wave, resulting in the rippling modulation as found in Fig.~\ref{fig1-hybrid}(e).

	To gain more insight into the daughter AW, we probe the AW signatures versus time at a fixed location in front of the injection point and further separate the forward/backward waves using the modified Elsasser variables
	\begin{equation}
		Z^{\pm}=\frac{1}{2v_A}(\delta v\cdot R_{bv}\mp \frac{\delta B}{\sqrt{\mu_0\rho}}),
		\label{eq:elsasser}
	\end{equation}
	where $Z^+$ and $Z^-$ refer to the forward wave and the backward wave, respectively. For the present case, $R_{bv}\simeq 0.78$ and the difference between the magnetic and velocity field components can be clearly seen by comparing Figs.~\ref{fig1-hybrid}(b) and (c). 
	The separated waves and corresponding Fourier spectra are presented in Figs.~\ref{fig1-hybrid}(g) and (h). Strong bursts of backward daughter AW (red color) due to the PDI are seen at later times ($t\gtrapprox 100T_0$). Its spectrum shows a prominent peak around $0.96\omega_0$, consistent with the anticipated central frequency of a lower sideband at $\omega_-/\omega_0=1-2\sqrt{\beta}\simeq 0.96$. A broad bandwidth also occurs, inherent to the finite instability bandwidth at high wave amplitudes~\cite{li2022parametric} and also likely enhanced by a spread in $\beta$ due to nonlinear modification of the plasma density and temperature.

	\section{Threshold wave amplitude due to the balance between instability growth and Landau damping} \label{amplitude}

	The above example shows, with a simplified plane-wave setup, what the PDI may look like if it can indeed be excited in LAPD-like plasmas. 
	Next, using the same setup, we extend the study to a large parameter space. We group the simulations by two parameters, $\beta$ and $\delta B/B_0$, which are most relevant in characterizing the PDI. The results are summarized in Fig.~\ref{fig2-amp}(a), where typical LAPD conditions correspond to the bottom-left corner (outlined by the dashed square), i.e., $\beta\sim 10^{-4}-10^{-3}$ and $\delta B/B_0\sim 10^{-3}-10^{-2}$. Three ion-to-electron temperature ratios, $T_i/T_e=0.1, 0.25, 1$, are considered and marked in the plot as different colors. Each simulation is represented by a dot in the figure, where open dots refer to simulations with no PDI and filled dots indicate those with PDI. The criteria for PDI onset are based on both density and AW signatures, i.e., whether the peak amplitude of the lower Alfv\'enic sideband is greater than 5\% of the main peak at the fundamental frequency, and the averaged density fluctuation $\langle\delta\rho/\rho_0\rangle$ is greater than 5\%. We have also checked that, for those open-dot cases, no PDI is found even if the AW is injected over $1000T_0$. 
	
	It is seen from Fig.~\ref{fig2-amp}(a) that the AW amplitude must exceed a threshold to trigger PDI in the system and the threshold amplitudes increase with $\beta$ and $T_i/T_e$. In general, parametric instabilities occur when the pump amplitude exceeds a threshold determined by damping effects, and both the AW waves and acoustic waves may experience damping. In LAPD experiments, AW spatial damping is due to electron-ion collisions and/or AW Landau damping; both may be varied by scanning parameters such as the electron temperature~\cite{thuecks2009tests}. In the present simulations, the spatial damping of pump AWs due to electron-ion collisions is characterized by the finite $\eta$. While the resultant AW spatial damping is found to cause slightly weaker PDI development (compared to cases without spatial damping, i.e., $\eta\to 0$), its effect is negligible. Meanwhile, the acoustic modes in LAPD may be damped by ion-neutral collisions or collisionless Landau damping; only the latter is modelled in the current code. Therefore, the threshold AW amplitudes in the present investigation should be determined by the balance between the PDI growth and Landau damping. 
	
	The PDI growth rate may be approximated in the low-beta, low-amplitude limit~\cite{sagdeev1969nonlinear} as
	\begin{equation}
		\gamma_g/\omega_0\simeq 0.5(\delta B/B_0)\beta^{-1/4}.
		\label{eq:pdi-growth}
	\end{equation}
	It should be emphasized that this normalized growth rate was obtained in the low-frequency single-fluid limit ($\omega_0/\Omega_{ci}\to 0$), but it is found to approximate our finite-frequency ($\omega_0/\Omega_{ci}\sim0.6$) cases well; we have checked against two-fluid growth rates~\cite{hollweg1994beat} and found that noticeable differences start to appear only when $\delta B/B_0>10^{-2}$ for typical LAPD beta values. 
	On the other hand, we take the Landau damping rate, $\gamma_d/\omega\simeq \sqrt{T_i/T_e}$, for the Cauchy ion distribution~\cite{gurnett2005introduction}, where $\omega$ is the acoustic wave frequency. Using the Cauchy distribution greatly simplifies the calculation of Landau damping and the Cauchy distribution reasonably approximates the actual Gaussian distribution in the ion velocity range of $|v_i/v_{th,i}|<1.5$, where $v_{th,i}=\sqrt{k_BT_i/M_i}$ is the ion thermal speed with $M_i$ being the ion mass and $k_B$ the Boltzmann constant (though the Cauchy distribution has a heavier tail at larger $v_i$). These simplifications are justified as we aim at providing a physical interpretation for the threshold-amplitude behavior, instead of an exact numerical prediction; the latter would require considering effects neglected by the simplifying assumptions made in Sec.~\ref{setup}. In the low-beta regime of PDI, the acoustic wave frequency satisfies $\omega/\omega_0\simeq 2\sqrt{\beta}$. Cast in units of $\omega_0$, the Landau damping rate takes the form
	\begin{equation}
		\gamma_d/\omega_0\simeq 2\sqrt{\beta}\sqrt{T_i/T_e}.
		\label{eq:landau-damping}
	\end{equation}
	In order to trigger PDI, one needs to satisfy $\gamma_g>\gamma_d$, which is translated into a threshold for the AW amplitude in a collisionless plasma
	\begin{equation}
		\delta B/B_0>4\beta^{3/4}\sqrt{T_i/T_e}.
		\label{eq:threshold-amp}
	\end{equation}
	This formula is appended as color lines in Fig.~\ref{fig2-amp}(a) for different $T_i/T_e$, which show reasonable agreement with the simulation results, especially in the LAPD parameter regime. The agreement confirms that the threshold is determined by the competition between PDI growth and Landau damping in our simulations. Importantly, these curves cross the bottom-left corner, an indication that the excitation of PDI might be within reach at LAPD if effects neglected by our model (see discussion in Sec.~\ref{discussions}) do not significantly modify the balance between growth and damping. 
	
	\begin{figure*}[t]
		\centering
		\includegraphics[width=0.9\textwidth]{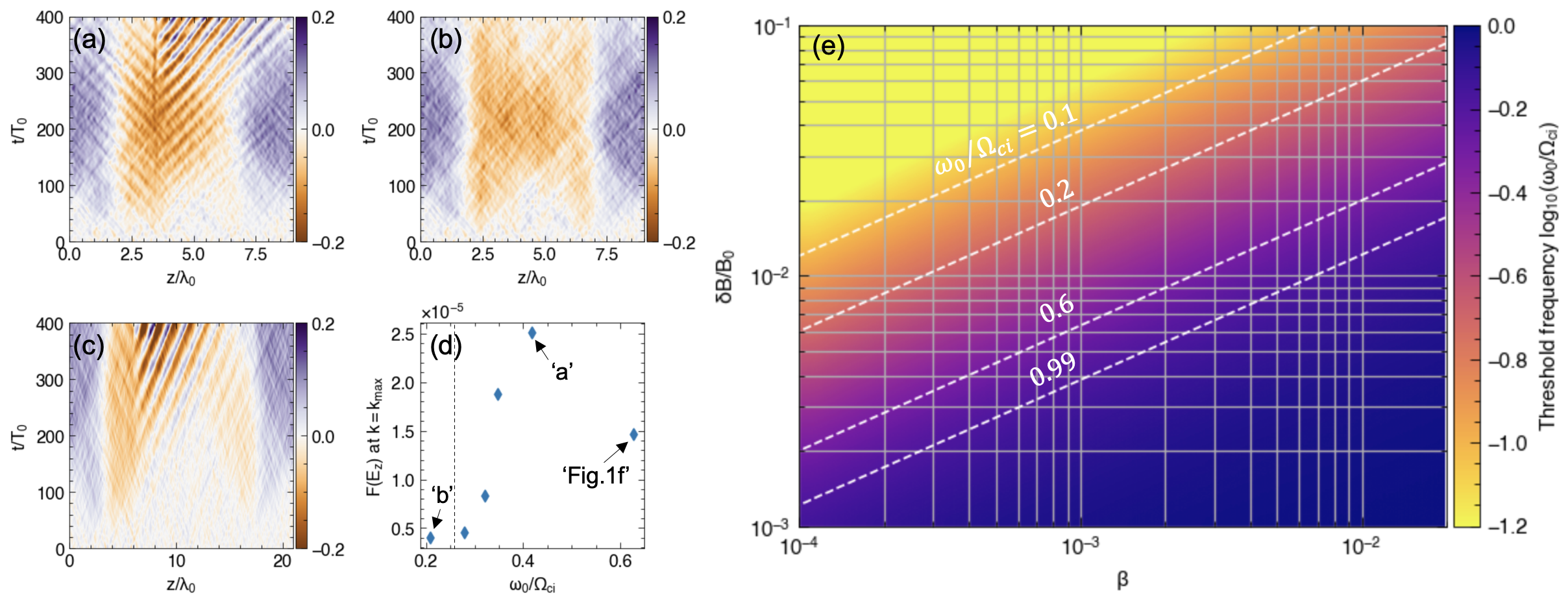}
		\caption{Threshold Alfv\'en wave frequency for triggering the parametric decay instability. (a, b) Spatiotemporal evolutions of the density fluctuation for case $\omega_0/\Omega_{ci}=0.42, 0.21$, respectively. The other parameters are the same as for Fig.~\ref{fig1-hybrid}. (c) The same as (b) but for an enlarged simulation domain, three times bigger in $L$. (d) Maximum of the Fourier spectral amplitude of $E_z$ fluctuation for different $\omega_0/\Omega_{ci}$, where the vertical dashed line marks the threshold frequency predicted by Eq.~(\ref{eq:threshold-freq}). (e) Theoretical estimate of the threshold wave frequency over the full parameter space spanned by wave amplitude and plasma beta [i.e., Eq.~(\ref{eq:threshold-freq})], where $L=25d_i$ and the appended lines represent contours at specific $\omega_0/\Omega_{ci}$.}
		\label{fig3-freq}
	\end{figure*}
	
	A closer comparison between Eq.~(\ref{eq:threshold-amp}) and the simulation data in Fig.~\ref{fig2-amp} shows a few prominent differences (notice the log-scale axes). Firstly, the simulations show lower threshold amplitudes, especially when $T_i/T_e$ is small. Physically, $T_e$ determines the phase speed of acoustic wave and $T_i$ determines the ion thermal speeds. When $T_i/T_e\ll 1$, the phase speed is much larger than the thermal speed, and the Cauchy distribution used for Eq.~(\ref{eq:landau-damping}) tends to overestimate the damping rate because of a heavier tail in the velocity distribution. Therefore, the theoretical curves tend to overestimate the required AW amplitude when $T_i/T_e\ll 1$. Secondly, the simulations also show threshold amplitudes below the theoretical curves in the higher-beta, higher-amplitude regime, i.e., the upper-right corner of Fig.~\ref{fig2-amp}(a). In this regime, despite that Eq.~(\ref{eq:pdi-growth}) tends to slightly overestimate the growth rate, the much stronger acoustic mode damping causes significant plasma heating which may greatly reduce Landau damping rate. To see this clearly, we show the plasma heating results of two example cases in Figs.~\ref{fig2-amp}(b) and (c), which are marked in Fig.~\ref{fig2-amp}(a) as orange/green crosses, respectively. The orange-cross case ($\beta=5\times 10^{-4}, \delta B/B_0=1\times 10^{-2}, T_i/T_e=0.25$) sits above threshold curve but falls in the low-beta, low-amplitude regime. In this case, the ion phase phase-space [Fig.~\ref{fig2-amp}(b1)] and ion parallel temperature distributions [Fig.~\ref{fig2-amp}(b3)] are largely regular, and the modification to the ion distribution function [Fig.~\ref{fig2-amp}(b2)] is negligible, meaning Eq.~\ref{eq:landau-damping} is still valid. In contrast, the green-cross case ($\beta=3\times 10^{-3}, \delta B/B_0=5\times 10^{-2}, T_i/T_e=1$) has significantly stronger heating [Fig.~\ref{fig2-amp}(c3)]; see also the distortions found in the phase-space [Fig.~\ref{fig2-amp}(c1)] and velocity distributions [Fig.~\ref{fig2-amp}(c2)].

	\section{Threshold wave frequency due to instability development in LAPD-like bounded plasma} \label{frequency}

	Aside from the requirement on AW amplitude, another important parameter for consideration in experiments is what wave frequencies one should use. This is particularly relevant to the LAPD plasma which is bounded and normally contains a few AW wavelengths~\cite{gigliotti2009generation}. 
	It was argued~\cite{pesme1973parametric} that the absolute instability develops in a bounded unstable region if the size of the system, $L$, satisfies 
	\begin{equation}
		\gamma_g L/|V_1V_2|^{1/2}>\pi/2, 
		\label{eq:threshold-freq-original}
	\end{equation}
	where $V_1, V_2$ are the group velocity of the two daughter waves, respectively. Physically, this means the system must be large enough for the instability to emerge before the daughter waves convect away. In terms of the PDI investigated here, one has $V_1=\frac{d\omega_0}{dk_\parallel}$ and $V_2=c_s$, where $c_s$ is the acoustic wave speed. Making use of the AW dispersion relation $k_\parallel d_i=\frac{\omega_0/\Omega_{ci}}{\sqrt{1-(\omega_0/\Omega_{ci})^2}}$, one has $\frac{d\omega_0}{dk_\parallel}=\frac{\sqrt{1-(\omega_0/\Omega_{ci})^2}}{1+k_\parallel^2 d_i^2}v_A=[1-(\omega_0/\Omega_{ci})^2]^{3/2}v_A$. The acoustic wave speed takes the form $c_s=\sqrt{\frac{\gamma_ek_BT_e+\gamma_i k_B T_i}{M_i}}=\sqrt{5\beta/6}v_A$ where $\gamma_e=\gamma_i=5/3$ are the electron/ion polytropic indices and we have used $\beta=\beta_e+\beta_i=(1+T_e/T_i)\beta_i$ and $\beta_i=2(v_{th,i}/v_A)^2$. Notice that the form of $c_s$ is independent of the temperature ratio $T_i/T_e$ for a fixed $\beta$. Substituting the expressions for $V_1, V_2$ into Eq.~(\ref{eq:threshold-freq-original}), the requirement on the system size is translated to a threshold for the wave frequency 
	\begin{equation}
		F(\omega_0/\Omega_{ci})\equiv\frac{\omega_0/\Omega_{ci}}{[1-(\omega_0/\Omega_{ci})^2]^{3/4}}\gtrapprox \frac{\pi(5/6)^{1/4}}{L/d_i}\frac{\sqrt{\beta}}{\delta B/B_0},
		\label{eq:threshold-freq}
	\end{equation}
	where we have used the identity $c/v_A=\omega_{pi}/\Omega_{ci}$ and $F(\omega_0/\Omega_{ci})$ is a monotonic increasing function of $\omega_0/\Omega_{ci}$. The result is independent of $T_i/T_e$ because the growth rate has no $T_i/T_e$ dependence [i.e., Eq.~(\ref{eq:pdi-growth})], which is different from the requirement on wave amplitude. Using $\omega_0=2\pi v_A\sqrt{1-(\omega_0/\Omega_{ci})^2}/\lambda_\parallel$ (with $\lambda_\parallel=2\pi/k_\parallel$), the frequency threshold can be recast as 
	\begin{equation}
		\frac{L}{\lambda_\parallel}\gtrapprox 0.48\frac{\sqrt{\beta}}{\delta B/B_0}[1-(\omega_0/\Omega_{ci})^2]^{1/4},
		\label{eq:threshold-freq-transform}
	\end{equation}
	which demands sufficient number of wavelengths to be present in the plasma region. 
	
	To verify the threshold frequency requirement, we have performed a series of simulations with $\beta=5\times 10^{-4}, \delta B/B_0=1\times 10^{-2}, T_i/T_e=0.25$, the same as the case of Fig.~\ref{fig1-hybrid}, but having different $\omega_0/\Omega_{ci}$. The case of Fig.~\ref{fig1-hybrid}(f) corresponds to $\omega_0/\Omega_{ci}\sim 0.63$ and we also select two other cases of $\omega_0/\Omega_{ci}\sim 0.42, 0.21$ and present their space-time density fluctuations in Figs.~\ref{fig3-freq}(a) and (b), respectively. It indeed shows that, as the pump frequency drops, the density fluctuations become weakened and eventually disappear for the case $\omega_0/\Omega_{ci}\sim 0.21$. Note that, for the present wave and plasma parameters, the threshold frequency predicted from Eq.~(\ref{eq:threshold-freq}) is $\omega_0/\Omega_{ci}>0.26$, consistent with the above simulations. The agreement is better illustrated by Fig.~\ref{fig3-freq}(d), which plots the maximum strength of acoustic wave (in terms of longitudinal $E_z$ field) for a series of $\omega_0/\Omega_{ci}$. A clear threshold behavior around $\omega_0/\Omega_{ci}\sim 0.3$ is seen, below which the wave strength is at the noise level and above which a sharp increase in the wave strength is found. 
	
	For the case with lowest frequency $\omega_0/\Omega_{ci}\sim 0.21$, we find in simulation $L/\lambda_\parallel\sim 0.83$, meaning the plasma length is shorter than one Alfv\'en wavelength.  Eq.~(\ref{eq:threshold-freq-transform}) demands $L/\lambda_\parallel>1.1$ for the parameters used, in order to see PDI. To verify Eq.~(\ref{eq:threshold-freq-transform}), we keep the same low frequency $\omega_0/\Omega_{ci}\sim 0.21$ but make $L$ three times bigger, i.e., $L/\lambda_\parallel\sim 2.5$. As shown in Fig.~\ref{fig3-freq}(c), the density fluctuations indeed reappear in the larger system. 
	
	Using Eq.~(\ref{eq:threshold-freq}) and $L=25 d_i$, the dependence of threshold frequency on the broad parameter space spanned by $\beta$ and $\delta B/B_0$ is shown in Fig.~\ref{fig3-freq}(e). The contours show the dependencies for a few constant $\omega_0/\Omega_{ci}$. Generally, the onset of PDI demands higher threshold frequencies towards the lower-right corner of the parameter space and vice versa. Because the threshold frequency in Eq.~(\ref{eq:threshold-freq}) is inversely proportional to the wave amplitude, the threshold frequency increases quickly with $\beta$ at small wave amplitudes. One should avoid approaching $\omega_0/\Omega_{ci}=1$ (bottom-right corner) where strong ion cyclotron resonances for a left-hand wave will be invoked. This plot would be useful for guiding the choice of wave frequency in future experiments.

	\section{Discussions}  \label{discussions}
	
	We have so far considered two critical parameters, AW amplitude $\delta B/B_0$ and frequency $\omega_0/\Omega_{ci}$, and studied their dependencies on plasma beta, temperature, and size to excite PDI in a bounded plasma with LAPD-like wave injection. However, there are several other factors that need to be considered for optimizing PDI experiments.
	
	Firstly, in presenting the threshold $\delta B/B_0$ versus $\beta$ (i.e., Fig.~\ref{fig2-amp}), the background field $B_0$ is involved in both $\delta B/B_0$ and $\beta=n_0k_BT/(B_0^2/2\mu_0)$, rendering $B_0$ a new degree of freedom. How to choose $B_0$ is of practical importance in LAPD experiments. To clarify this point, we introduce the parameter $\xi$ which is defined as the ratio between PDI growth and Landau damping, i.e., 
	\begin{equation}
		\xi=\frac{\gamma_g}{\gamma_d}=\frac{1}{4}\frac{\delta B}{B_0}\frac{(B_0^2/2\mu_0)^{3/4}}{(n_0KT_e+n_0KT_i)^{3/4}}\sqrt{\frac{T_e}{T_e}}\propto \sqrt{B_0}.
	\end{equation}
	The $\sqrt{B_0}$ dependence suggests that PDI favors higher-$B_0$ regime, which corresponds to the lower-beta, lower-normalized-amplitude regime.

	Secondly, the ion-to-electron temperature ratio has been shown as an important parameter [Fig.~\ref{fig2-amp}(a)]. Recently, an upgrade was made in LAPD including the installation of a new $\rm LaB_6$ cathode, which increased the accessible temperature ranges for both electrons (1 to 15 eV) and ions (sub-eV to 10 eV) and may also indicate more flexibility in the temperature ratios. More importantly, the high electron temperatures will allow us to achieve higher pump AW amplitudes by minimizing both electron-ion collisional damping and AW Landau damping~\cite{thuecks2009tests}. 
	
	Thirdly, actual LAPD experiments usually involve a finite wave scale in the transverse direction, which is determined by the antenna size to be used~\cite{gigliotti2009generation}. The source size is typically much smaller than $\lambda_\parallel$, but the wave remains collimated over the relatively small plasma dimension due to insignificant geometric attenuation of AWs~\cite{karavaev2011generation}, which may justify the use of the quasi-1D setup in our work. However, the finite $k_\perp\rho_s \sim 0.2$ (with $\rho_s$ being the ion sound gyroradius) arising from the source size could be an important factor that may potentially complicate the PDI physics~\cite{dorfman2016observation}. A detailed theoretical understanding of the finite $k_\perp$ effect on PDI has not yet been developed.
	
	Finally, the ion-neutral collisions present in the LAPD may enhance the damping of the acoustic wave, raising the threshold curves in Fig.~\ref{fig2-amp}(a), i.e., requiring higher AW amplitudes.  We are currently implementing 3D wave injection of finite source sizes and ion-neutral collisions in the H3D code. Studies of PDI with these additional features will be presented in a separate  work.

	\section{Summary}
	
	In summary, we have presented novel hybrid simulation capabilities to model continuous wave injection in a bounded plasma with absorption wave boundaries and realistic wave-plasma parameters. Such simulations could be a powerful tool to investigate a variety of LAPD AW experiments such as beat wave excitation~\cite{dorfman2013nonlinear} and turbulence generation~\cite{howes2012toward}. Here, we have focused on the PDI of nonlinear AWs, a fundamental process in a magnetized plasma. Using parameters that resemble LAPD conditions, our simulations and accompanied theoretical analyses have provided important insight into the requirements for exciting PDI in a LAPD-like plasma, including the threshold values for AW amplitude and frequency. Other factors not included in the present simplified model (plane wave injection, collisionless plasma) are also briefly discussed. The current work is a critical step toward direct numerical modeling of LAPD plasmas and it will guide future laboratory demonstration of the fundamental PDI process which could have significant implications in space and astrophysical plasmas.

	\section{Acknowledgement}
	This work is supported by a National Science Foundation and Department of Energy Partnership in Basic Plasma Science and Engineering program under the grant DE-SC0021237. X.F. is also supported by the Los Alamos National Laboratory/Laboratory Directed Research and Development program and DOE/Office of Fusion Energy Sciences. S.D. is also supported by the National Aeronautics and Space Administration (NASA) grant 80NSSC18K1235. F.L. and X.F. acknowledge the Texas Advanced Computing Center (TACC) at The University of Texas at Austin and the National Energy Research Scientific Computing Center (NERSC) for providing HPC and visualization resources. NERSC is supported by the Office of Science of the U.S. Department of Energy under Contract No. DE-AC02-05CH11231.
	
	\section{Data availability statement}
	The data that support the findings of this study are available from the corresponding author upon reasonable request.
	
	\bibliography{ref_pdi}
	
\end{document}